# Evidence for spin-flip scattering and local moments in dilute fluorinated graphene

X. Hong[1,3], K. Zou[1], B. Wang[1], S.-H. Cheng[1] and J. Zhu[1,2*]

1. Department of Physics, The Pennsylvania State University, University Park, Pennsylvania, 16802, USA
2. The Materials Research Institute, The Pennsylvania State University, University Park, Pennsylvania, 16802, USA
3. Department of Physics and Astronomy and Nebraska Center for Materials and Nanoscience, University of Nebraska-Lincoln, Lincoln, Nebraska 68588

*Email: jzhu@phys.psu.edu

Submission Date: May 10, 2011

**Abstract:**

The issue of whether local magnetic moments can be formed by introducing adatoms into graphene is of intense research interest because it opens the window to fundamental studies of magnetism in graphene, as well as of its potential spintronics applications. To investigate this question we measure, by exploiting the well-established weak localization physics, the phase coherence length $L_\Phi$ in dilute fluorinated graphene. $L_\Phi$ reveals an unusual saturation below ~ 10 K, which cannot be explained by non-magnetic origins. The corresponding phase breaking rate increases with decreasing carrier density and increases with increasing fluorine density. These results provide strong evidence for spin-flip scattering and points to the existence of adatom-induced local magnetic moment in fluorinated graphene. Our results will stimulate further investigations of magnetism and spintronics applications in adatom-engineered graphene.

Adatom-engineering has become an active research front in the field of graphene recently because it is a powerful tool to alter, control and engineer the transport, optical, and potentially magnetic properties of graphene. Examples include the introduction of a mobility edge [1-4], and the opening of a band gap and fluorescence [5-9] in hydrogenated, oxygenated and fluorinated graphene. The atomic size of adatoms, their chemical bonding with graphene, and the unique dispersion of the graphene electrons make adatoms a unique type of defects that interact strongly with electrons in graphene [10, 11]. This interaction may offer an effective way to induce local magnetic moments into non-magnetic graphene and lead to the exploration of various magnetic ground states in this unique two-dimensional system [12]. Magnetic graphene is also of significant technological interest in spintronics applications. In stark contrast to the large number of theoretical studies, the experimental evidence on adatom-induced magnetism in monolayer graphene remains elusive [12]. Recently, we observed a colossal negative magnetoresistance in dilute fluorinated graphene (DFG), suggesting the involvement of magnetic moments [4]. Here, we report an anomalous saturation of the phase coherence length at low temperatures in the same system. This saturation cannot be accounted for by conventional mechanisms based on sample size or magnetic contaminations, but instead points to spin-flip scattering caused by adatom-induced local magnetic moment. The spin-flip rate is tunable by electron density and fluorine density. These observations suggest that functionalizing graphene with fluorine can potentially offer a gate-controllable approach to manipulate spins in graphene, which is essential to the operation of spintronics devices.

Dilute fluorinated graphene samples, where the fluorine adatom covalently attaches to the graphene plane (inset of Fig. 1(a)), are prepared in $CF_4$ plasma as described in Ref. [4]. The plasma process is clean and reversible, producing negligible amount of vacancies [4]. We estimate the F-adatom density $n_F$ to be 0.5, 2.2, and $2.4 \times 10^{12}$/cm$^2$ respectively in samples A-C using the intensity ratio of the $D$ and $G$ peaks in their Raman spectra, following the empirical relation $I_D/I_G = 1.02 \, n_{\text{def}}$ ($10^{12}$/cm$^2$) obtained for atomic defects in Refs. [13, 14]. The F-adatom density $n_F$ in sample B is also independently determined to be $2 \times 10^{12}$/cm$^2$ by scanned tunneling microscopy measurements in Ref. [4]. These densities correspond to a very dilute F:C ratio of 1 to several thousands. DFG samples are fabricated into field effect transistor (FET) devices using standard lithography and measured in a perpendicular magnetic field with low-frequency lock-in techniques. A low excitation current of 0.5 to 50 nA is used to avoid Joule heating [4].

Figures 1(a) and 1(b) plot the zero-field sheet conductance σ of samples A and B as a function of the backgate voltage $V_g$ at different temperatures (5-200 K). Sample C exhibits $\sigma(V_g)$ similar to that of sample B [14]. Although the high-temperature $\sigma(V_g)$ resembles that of pristine graphene [15], the Hall mobility $\mu_{\text{Hall}} = \sigma/ne$ is significantly lower in DFG. At 200 K and $n = 3 \times 10^{12}$/cm$^2$, $\mu_{\text{Hall}} = 1,000$ cm$^2$/Vs in sample A and 310 cm$^2$/Vs in samples B and C, in contrast to $\mu_{\text{Hall}} > 10,000$ cm$^2$/Vs in our un-treated graphene samples, where charged impurity scattering dominates [15, 16]. At the single impurity level, atomic defects are shown to scatter electrons in graphene strongly following the midgap state scattering model (Eq. (1)), where the conductivity is inversely proportional to $n_F$ [10, 11, 17, 18].

$$\sigma^{midgap} = \frac{e^2}{h} \frac{2}{\pi} \frac{n}{n_F} \ln^2(k_F R_0) \qquad (1).$$

Here, $R_0$ is the interaction radius of the defect potential and $k_F$ the Fermi wave-vector. The mobility ratio between samples A and B leads to $n_F \sim 6 \times 10^{11}$/cm$^2$ in sample A, in agreement with $n_F \sim 5 \times 10^{11}$/cm$^2$ estimated from Raman data. In Fig. 1(a), Eq. (1) is plotted as dashed curves in sample A using $n_F = 5$-$6 \times 10^{11}$/cm$^2$ and $R_0 = 4.4$-$4.0$ Å. It describes the $\sigma(V_g)$ curve very well at high temperature. As $T$ decreases and/or $n_F$ increases, localization effects become important [4] and Eq. (1) no longer applies. As Fig. 1(b) and Ref. [14] show, localization effects are more pronounced in samples B and C, where the shape of $\sigma(V_g)$ can no longer be described by Eq. (1).

Previously, we observed a carrier-density driven crossover from weak localization ($\sigma > 2e^2/h$) to two-dimensional (2D) variable-range hopping ($\sigma < 2e^2/h$) in the temperature dependence of $\sigma$ in DFG samples. In the hopping regime, DFG samples exhibit very large negative magnetoresistance. One possible explanation involves local magnetic moments induced by F-adatoms [4]. Such local moments can be sensitively probed by phase breaking measurements, where magnetic impurities break the time-reversal symmetry of coherent backscattering paths [19] and contribute to phase decoherence in addition to conventional phase-breaking mechanisms in 2D such as electron-electron scattering and electron-phonon scattering [20-23]. In graphene, the theory that relates the phase coherence length $L_\Phi$ to the magnetoconductance $\sigma(B)$ is given by Eq. (2) [24]:

$$\frac{\pi h}{e^2} \Delta\sigma(B) = F\left(\frac{4l_B^{-2}}{L_\Phi^{-2}}\right) - F\left(\frac{4l_B^{-2}}{L_\Phi^{-2} + 2L_i^{-2}}\right) - 2F\left(\frac{4l_B^{-2}}{L_\Phi^{-2} + L_i^{-2} + L_*^{-2}}\right) \qquad (2).$$

Here, the first term corresponds to the conventional weak localization including all phase breaking processes. The second and third terms describe the anti-localization effect unique to graphene due to intervalley ($L_i$) and chirality-breaking intravalley ($L_*$) scatterings. While atomically sharp defects cause both scattering, $L_*$ is also affected by scattering with ripples, dislocations and the warped Fermi surface [24, 25]. $l_B = (\hbar/eB)^{1/2}$ is the magnetic length and $F(z) = \ln z + \Gamma(0.5 + z^{-1})$, where $\Gamma$ is the digamma function. Equation (2) provides a very good description of magnetoconductance in pristine graphene, where the effects of both localization and anti-localization are observed [26-28].

We study the phase breaking processes in DFG samples following Eq. (2). Only high carrier densities in the weak localization regime ($\sigma > 2e^2/h$) are analyzed to avoid complications associated with strong localization [29]. Figure 2(a) shows $\sigma(B)$ of sample B at $n = 4.2\times10^{12}/cm^2$ and varying temperatures. Fits to Eq. (2) are shown as dashed lines and provide an excellent description of data at all temperatures, from which we extract $L_\Phi(T)$, $L_i(T)$ and $L_*(T)$. To simplify the high-field fitting, we have assumed $L_i \approx L_*$ in our samples. This assumption does not affect the determination of $L_\Phi$, which is solely determined by the low-field data at $B < 1$ T. The extracted $L_i \approx L_* = $ 10-20 nm is much smaller than those reported in pristine graphene [26, 27], confirming the expected dominance of scattering with atomically sharp F-adatoms. $L_i$ and $L_*$ are also roughly temperature-independent. Good fits to Eq. (2) are obtained on all samples. In Fig. 2(c), we plot $\sigma(B)$ and fits at $n = 2.5\times10^{12}/cm^2$ on sample A. In this sample, Shubnikov de Hass oscillations start at roughly 4 T, indicating that apart from a dilute coverage of F-adatoms, the quality of the sample remains high and similar to pristine graphene. The occurrence of the magneto-oscillations at high field does not contradict the presence of magnetic moments since the magnetic field created by such low-density moments is negligible. Similar phenomenon has also been observed in Mn-doped InAs quantum wells [30].

It is interesting to point out that although $\sigma(B)$ appears to have saturated at high field in Fig. 2(a), a considerable $T$-dependence remains, as demonstrated by the temperature dependence of $\sigma(T)$ at 0 T and 8.8 T in Fig. 2(b). This $T$-dependence arises from the interplay of all three terms in Eq. (2), which is more complex than that of a conventional 2D electron gas [22].

Figure 3(a) plots the phase coherence length $L_\Phi$ vs. $T$ for several densities from 0.6 to $3.8\times10^{12}/cm^2$ in sample A. $L_\Phi$ increases with decreasing $T$ but saturates towards a constant at low temperature. The corresponding phase breaking rate $\tau_\Phi^{-1}$ is calculated by $\tau_\Phi^{-1} = D/L_\Phi^2$, where $D = v_F l/2$ is the diffusion constant determined by the mean free path $l = (\sigma/k_F)h/2e^2$ and $v_F = 1\times10^6$ m/s. Here, $\sigma$ (200 K) is used as the Drude conductivity. The resulting $\tau_\Phi^{-1}(T)$ is plotted in Fig. 3(b). Similar to $L_\Phi(T)$, $\tau_\Phi^{-1}(T)$ shows a clear tendency towards saturation at low temperature. Similar behavior is observed in all DFG samples.

The *T*-dependence of $\tau_\Phi^{-1}$ is well described by the following fitting equation:

$$\tau_\Phi^{-1} = aT + bT^2 + \tau_{sat}^{-1} = \alpha k_B T \ln g / \hbar g + bT^2 + \tau_{sat}^{-1}; \qquad (3),$$

where $g = \sigma/(e^2/h)$ is the normalized conductance. The linear *T* term is commonly attributed to electron-electron scattering in the diffusive regime of a 2D electron gas [20-22]. Our fits yield $a = 0.03$-$0.06$ ps$^{-1}$K$^{-1}$ or $\alpha = 1$-$2$, in good agreement with previous results in pristine graphene [26]. With a coefficient of $b < 2 \times 10^{-4}$ ps$^{-1}$K$^{-2}$, the $T^2$ term, also attributed to electron-electron scattering [27, 28], plays a minor role at low temperatures. Another *T*-dependent inelastic process, electron-phonon scattering in graphene is negligible in this temperature range [26-28]. In summary, the *T*-dependence of $\tau_\Phi^{-1}$ of the DFG samples at high temperature agrees with that of pristine graphene. This is not surprising since F-adatoms are not expected to modify electron-electron interactions.

What distinguishes DFG samples from pristine graphene is the anomalous saturation of $L_\Phi$ and $\tau_\Phi^{-1}$ at low temperature. In Figs. 3(a) and 3(b), fitting to Eq. (3) yields $L_{sat}$ in the range of 80 – 400 nm. This observation is very different from those in pristine graphene, where $L_\Phi$ saturates at several microns–at least 10 times larger than those of DFG samples at comparable densities and the sample boundary is believed to be responsible [26, 27]. This cannot be the case in our DFG samples since all sample dimensions are greater than 3 μm (Fig. 1(b) inset), which is about a factor of ten larger than the largest $L_{sat}$.

We have carried out similar measurements and analysis on a control sample A$_{def}$ to rule out the possibility of magnetic contaminations or vacancies generated in the plasma process giving rise to the saturation of $L_\Phi$. Sample A$_{def}$ is fluorinated under similar condition as sample A and then completely defluorinated using procedures described in Ref. [4]. Any magnetic contamination or vacancies present in sample A would be present in sample A$_{def}$ as well. Raman spectrum of sample A$_{def}$ shows a very small $I_D/I_G$ ratio of 1:14, indicating a negligible vacancy density. The inset of Fig. 3(b) plots $\tau_\Phi^{-1}$ vs. *T* in samples A and A$_{def}$ at the same density $n = 1.4 \times 10^{12}$/cm$^2$. In stark contrast to A, $\tau_\Phi^{-1}$ in A$_{def}$ does not saturate down to the lowest temperature. Fitting to Eq. (3) yields a lower bound of $\tau_{sat} \approx 100$ ps, which is 30 times longer than $\tau_{sat}$ in sample A at this density and comparable to those observed in pristine graphene [26, 27]. We thus conclude that the origin of the observed saturation in $L_\Phi$ must originate from the F-adatoms.

Furthermore, $\tau_{sat}$ appears to depend on both the carrier density $n$ and the fluorine density $n_F$. In Fig. 4, we plot $\tau_{sat}$ vs. $n$ in three DFG samples A-C. As $n$ increases from 0.6 to 3.8 $\times 10^{12}$/cm$^2$ in sample A, $\tau_{sat}$ increases from 1 to 14 ps. Both samples B and C have approximately 3-4 times higher fluorine coverage than sample A and the measured $\tau_{sat}$'s are 3-5 times lower at the same carrier densities. This correlation further supports F-adatoms as the source of the observed anomalous phase breaking. More studies with systematic tuning of the F-adatom and carrier densities are necessary to establish a quantitative understanding of $\tau_{sat}$ in DFG samples.

Our results are consistent with the presence of spin-flip scattering and point to the existence of adatom-induced local magnetic moments. Such scattering breaks the time reversal symmetry of forward and backward trajectories and leads to phase breaking. The spin-flip scenario can account for several features of our observations. The spin-flip rate $\tau_{sf}^{-1}$ is related to the Kondo effect and is give by the Nagaoka-Suhl formula [19, 31, 32]:

$$\frac{1}{\tau_{sf}} = \frac{n_{mag}}{\pi \hbar N(E_F)} \frac{\pi^2 S(S+1)}{\pi^2 S(S+1) + \ln^2\left(T/T_K\right)} \quad \text{with} \quad k_B T_K = A \exp\left(-1/N(E_F)J\right) \quad (4),$$

where $n_{mag}$ is the density of non-interacting magnetic impurities, $S$ the magnetic moment and $T_K$ the Kondo temperature. $A = 10$ eV is the cut-off energy and $J$ the Kondo exchange energy. $N(E_F)$ is the density of states at the Fermi level. Setting $n_{mag} = n_F$ and $\tau_{sat}^{-1} = \tau_{sf}^{-1}$, Eq. (4) naturally leads to $\tau_{sat} \sim 1/n_F$. Using $S = 1/2$ and approximating $N(E_F)$ with the density of states of graphene, we estimate the Kondo temperature $T_K$ for each carrier density in Fig. 3(b). $T_K$ increases with decreasing $n$, reaching approximately 0.01 mK at $n \sim 0.6 \times 10^{12}$/cm$^2$. This corresponds to an estimated exchange energy $J$ of 5 meV. In the measured temperature range $T \gg T_K$, $\tau_{sf}^{-1}(T)$ is nearly $T$-independent, which is consistent with a constant $\tau_{sat}^{-1}$ extracted from Eq. (3). Although a direct signature of the Kondo effect in the conductivity is inaccessible using current cryogenic technology, phase-breaking measurements can nonetheless reveal the presence of magnetic moments sensitively. Indeed, similar situations were demonstrated in metallic wires, where magnetic impurities in the extreme dilute limit of 0.01 ppm were identified [19].

The hypothesis of F-adatom induced magnetic moments naturally connects the anomalous phase saturation observed here with our previous report of colossal magnetoresistance in the variable-range hopping regime at yet lower carrier densities [4]. In Fig. 4, the decrease of $\tau_{sat}$ with decreasing $n$ suggests enhanced spin-flip scattering, or stronger magnetic interaction, at lower densities. A continuation of this trend is consistent with the proposed formation of magnetic polarons in the variable-range hopping regime [4]. First principle calculations have indicated that hydrogen adatoms induce a local moment [12], which has not been confirmed experimentally; the situation with fluorine adatoms is more debatable [33]. We expect that the intriguing behavior exhibited by the DFG samples will stimulate more in-depth studies on the transport and magnetic properties of adatom-engineered graphene.

Our observations point to the existence of local moments in dilute fluorinated graphene; the rate of the spin-flip scattering induced by F-adatoms appears to be tunable by both the fluorination level and the carrier density in a single device. Long spin diffusion length of several μm has been reported in graphene [34, 35]. However, because of a weak spin-orbit coupling, pristine graphene has no intrinsic mechanism to manipulate spin, which is essential to spintronics applications. This work shows that controlled fluorination, combined with a gate-tunable carrier density can potentially operate as a spin-FET. Thus DFG spin-FETs connected by graphene interconnects may represent a new paradigm of spin circuits and open new avenues of research in all-carbon spintronics.

In conclusion, we observe an anomalous saturation of the phase coherence length in dilute fluorinated graphene. This saturation provides evidence for spin-flip scattering, which points to the presence of adatom-induced local magnetic moments. The spin-flip rate is tunable via carrier density and fluorine coverage. Functionalizing graphene with fluorine may offer a platform for studying magnetism in this unusual two-dimensional electron gas and a gate controllable, lithography-compatible approach to control spins in graphene spintronics devices.

We thank B. Altshuler, V. Falko, J. Sofo, and J. Jain for helpful discussions. This work is supported by NSF Grants CAREER No. DMR-0748604, MRSEC No. DMR-0820404 and ONR Grant No. N00014-11-1-0730. X.H. acknowledges support from NSF MRSEC Grant No. DMR-0820521 and Nebraska EPSCoR First Award. The authors acknowledge use of facilities at the PSU site of NSF NNIN.

**Figure Captions:**

Fig. 1 $\sigma(V_g)$ of (a) sample A and (b) sample B at $T = 5, 20, 50, 100, 200$ K (bottom to top). A charge neutrality point offset of 15 V (A) and 16 V (B) has been subtracted from $V_g$ respectively. The dashed line in (a) plots the midgap state scattering model (Eq. (1)) with $n_F = 6 \times 10^{11}/cm^2$ and $R_0 = 4.0$ Å. An equally good fit can be obtained using $n_F = 5 \times 10^{11}/cm^2$ and $R_0 = 4.4$ Å. Equation (1) applies to $|V| > V_{loc}$, where $V_{loc}$ corresponds to the Fermi energy $E_F = E_{loc} = \frac{\hbar v_F}{R_1 \sqrt{|\ln(R_0/R_1)|}}$, and $R_1$ is the fluorine-fluorine distance [10]. Insets to (a): A covalently bonded F-adatom; and (b): optical image of a typical DFG device.

Fig. 2 (a) Magnetoconductance of sample B at $n = 4.2 \times 10^{12}/cm^2$ with $T = 1.5$ K, 10 K, 20 K, 30 K, 50 K, 75 K, 100 K, 150 K, and 200 K (bottom to top). The dashed lines are fits to Eq. (2). (b) $\sigma(T)$ of sample B at $B = 0$ T (black square) and 8.8 T (red circle) and $n = 4.2 \times 10^{12}/cm^2$. The dashed lines indicate $\ln T$ temperature dependence. (c) Magnetoconductance of sample A at $n = 2.5 \times 10^{12}/cm^2$ and $T = 2.5$ K with fit to Eq. (2) (blue dashed line). The arrows mark the filling factors of the Shubnikov de Haas oscillations.

Fig. 3 (a) Phase coherence length $L_\Phi$ and (b) the corresponding phase breaking rate $\tau_\Phi^{-1}$ of sample A as a function of temperature at $n = 0.6, 1.1, 1.4, 2.5,$ and $3.8 \times 10^{12}/cm^2$ as indicated in the figure. Solid lines are fits to Eq. (3). The dashed lines in (b) correspond to the $aT + bT^2$ and the $\tau_{sat}^{-1}$ terms in Eq. (3), respectively, for $n = 3.8 \times 10^{12}/cm^2$. The intercept ranges from 3 to 15 K for the densities shown here. Inset: $\tau_\Phi^{-1}$ vs. $T$ for samples A (open symbol), and A$_{def}$ (solid symbol) at $n = 1.4 \times 10^{12}/cm^2$. The black solid line is a fit with only the $aT + bT^2$ terms in Eq. (3). Fits including $\tau_{sat} > 100$ ps fit data equally well (not shown).

Fig. 4 The saturated phase breaking time $\tau_{sat}$ as a function of carrier density for samples A-C. The dashed line serves as the guide to the eye.

Fig. 1

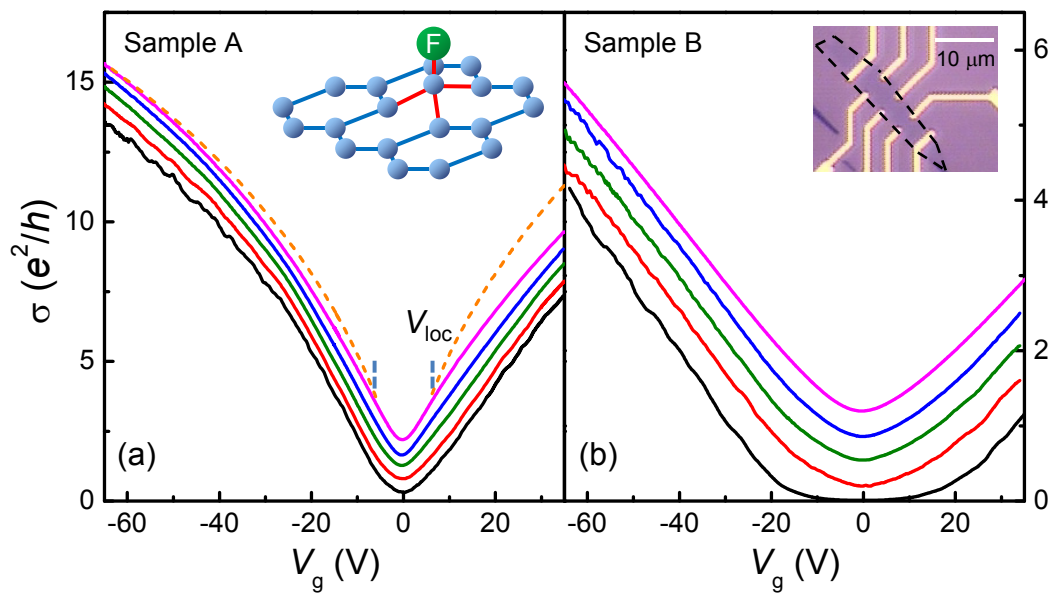

Fig. 2

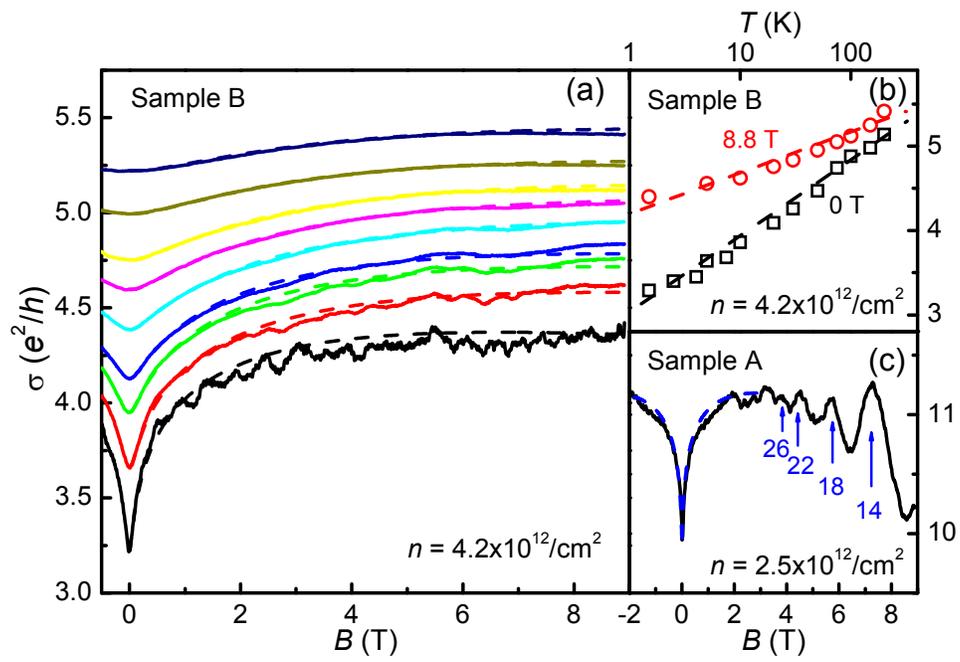

Fig. 3

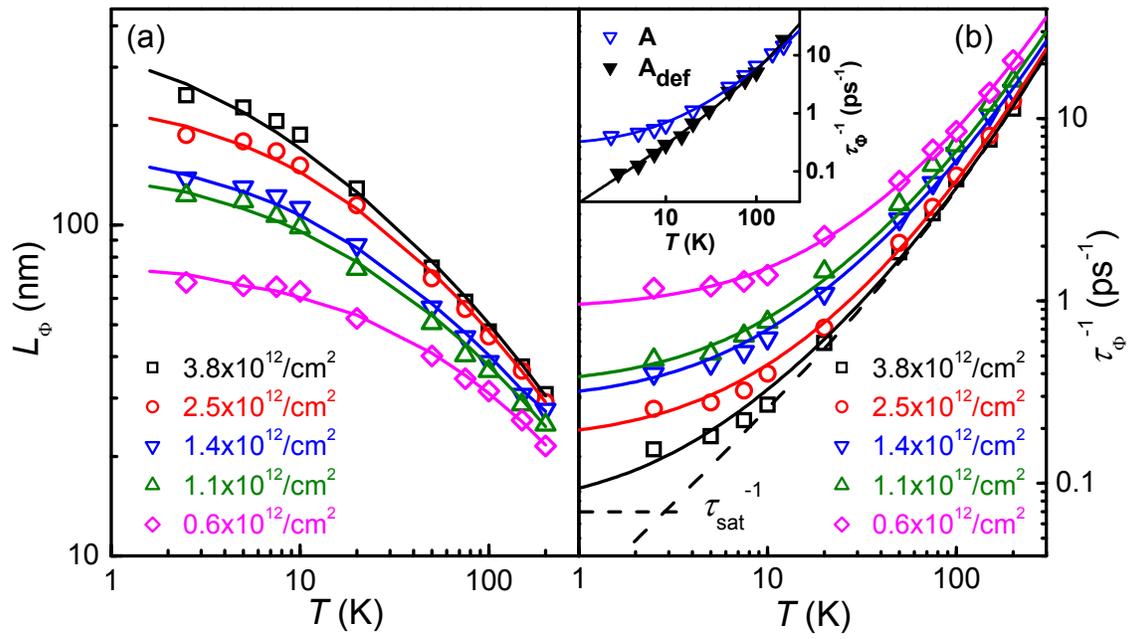

Fig. 4

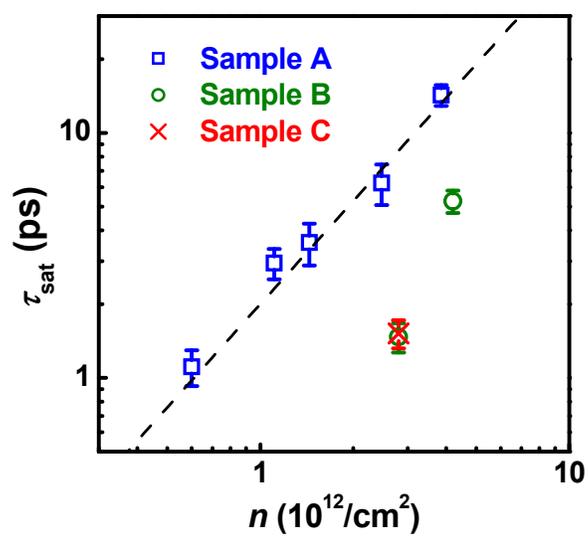

**Evidence for spin-flip scattering and local moments in dilute fluorinated graphene (Supplementary)**

X. Hong[1,3], K. Zou[1], B. Wang[1], S.-H. Cheng[1] and J. Zhu[1,2*]

1. Department of Physics, The Pennsylvania State University, University Park, Pennsylvania, 16802, USA
2. The Materials Research Institute, The Pennsylvania State University, University Park, Pennsylvania, 16802, USA
3. Department of Physics and Astronomy and Nebraska Center for Materials and Nanoscience, University of Nebraska-Lincoln, Lincoln, Nebraska 68588

*Email: jzhu@phys.psu.edu

**Online Supplementary Information Content:**

1 Raman characterization of the DFG samples

2 $\sigma(V_g)$ of Sample C

## 1 Raman characterization of the DFG samples

Figure S1 shows the Raman spectra of samples A, B and C. The intensity ratios between the D peak and G peak $I_D/I_G$ are 0.5, 2.2 and 2.5, respectively, in samples A-C. These correspond to $n_F = 0.5$, 2.2 and $2.4 \times 10^{12}/cm^2$ using the empirical relation $I_D/I_G \sim 102 \pm 2/L_D^2$, where $L_D$ is the average defect-defect distance, given by Lucchese *et al.* for dilute vacancy defects [1]. Our STM measurements on sample B gives $n_F = 2.0 \times 10^{12}/cm^2$ [2], in agreement with the above Raman estimate.

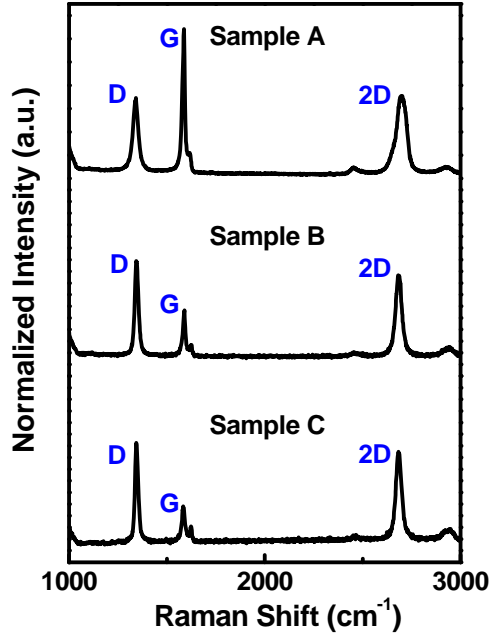

Fig. S1 Raman spectra for samples A, B, and C normalized to 2D intensity.

## 2 $\sigma(V_g)$ of Sample C

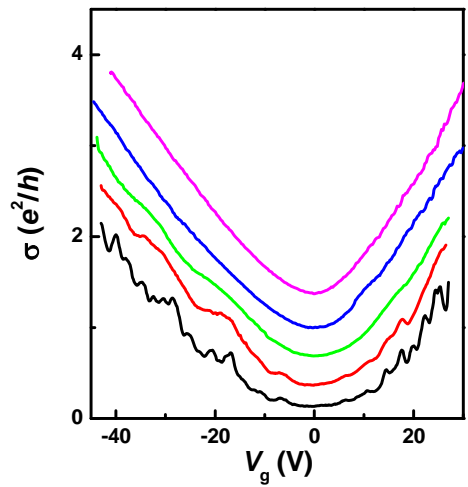

Fig. S2 Conductivity of sample C at 5 K, 20 K, 50 K, 100 K, and 200 K (bottom to top).

Figure S2 shows $\sigma(V_g)$ of sample C, which was fluorinated under similar condition as sample B. Its Hall mobility at $n = 3\times10^{12}/cm^2$ is 310 cm$^2$/Vs at 200 K, same as that in Sample B. The temperature and density dependence of $\sigma$ is also similar to Samples B.